\documentstyle[prl,aps,amssymb]{revtex}

\newcommand{\bra}[1]{\left\langle #1 \right|}
\newcommand{\ket}[1]{\left| #1 \right\rangle}
\newcommand{\pauli}{\hat{\sigma}}
\newcommand{\qE}[1]{\langle #1 \rangle}
\newcommand{\DecoKer}{K^N_{\rm d}}

\begin{document}
\twocolumn
\draft                          
\title{Phase space caustics in multi-component systems}
\author{Atushi Tanaka\cite{YRF}\cite{EMAIL}\cite{PresentAddress}}
\address{Department of Applied Physics, Tokyo Institute of Technology,
  Oookayama 2-12-1, Meguro 152, Tokyo, Japan \\
  and \\
  Yukawa Institute for Theoretical Physics,
  Kyoto University, Kyoto 606-01, Japan
  }

\date{January 8, 1998}

\maketitle
\begin{abstract}
  As examples of quantum-``classical'' coupling systems,
  multi-component systems are studied by semiclassical evaluations
  of the Feynman kernels in the coherent-state representation. From
  the observation of the phase space caustics due to the presence of
  the internal degree of freedom (IDF), two phenomena are explained in
  terms of the semiclassical theory: 
  (1) The quantum oscillations of the IDF induce quantum interference
  patterns in the Hushimi representation;
  (2) Chaotic dynamics destroys the coherence of the quantum oscillations.
\end{abstract}

%
Coupling a quantum system with a ``classical'' system provides not
only conceptual problems, e.g.\ the descriptions of measurement processes
only in terms of unitary time evolutions~\cite{dEspagnat}, but also
practical problems, e.g.\ interactions between electrons and nuclei in 
molecules~\cite{Molecule,Pechukas,StockThoss}. 
The difficulty that the classical concept is 
inapplicable for the ``classical subsystem'' in the coupled system
arises since the quantum and the classical subsystems become
entangled~\cite{Schroedinger} due to their interaction. 
However, it is possible to apply semiclassical methods, which elucidate
quantum dynamics in terms of classical dynamics, to quantum-classical
coupling systems.
It is natural to treat only the classical subsystem with semiclassical
methods for quantum-classical coupling systems, though it is formally
simple to apply a semiclassical method to the whole
system~\cite{StockThoss}.  

By applying a semiclassical theory only to the classical subsystem, this
letter elucidates the following two phenomena of the quantum-classical
coupling systems: 
The one is the interference phenomena of the classical subsystem due
to the coupling to the quantum subsystem; The other is the destruction of
the coherent quantum oscillation of the quantum subsystem by the
``chaotic'' dynamics of the classical subsystem. Note that
the notion chaos in quantum dynamics can be introduced only through
the semiclassical argument~\cite{Gutzwiller}. 

Throughout this letter, multi-component systems, e.g.\ electrons with a 
spin, and molecules that have quantized electrons, are employed as 
simple quantum-classical coupling systems. 
A quantum multi-component system consists of an internal and an
``external'' degrees of freedom (IDF and EDF, respectively): the IDF
is a quantum subsystem, i.e.\ the IDF is conveniently described by
matrices that have discrete indices rather than continuous indices; On
the other hand, the EDF can be regarded as a ``classical''
system. Namely, it is natural to employ continuous-valued variables
for describing the EDF.

\paragraph*{Interference phenomena of a ``classical'' subsystem due 
  to the coupling with a quantum system}
%
The emergence of the quantum interference phenomena implies the
complete breakdown of the classical picture. However, we can understand
the interference phenomena with a semiclassical argument. 
In the following, I will study the interference phenomena produced by a 
time evolution in the Feynman kernel in the coherent state
representation
\begin{equation}
  \label{eq:HeavyFK}
  K^{t}(q''p''\eta''; q'p'\eta') 
  \equiv \bra{q'' p'', \eta''} e^{-i \hat{H} t / \hbar }
                 \ket{q' p', \eta'},
\end{equation}
where \(\hat{H}\) is a Hamiltonian, \(\ket{q p}\) is 
an EDF's coherent state~\cite{Glauber}, which is a natural
correspondent of a point in the classical phase space, and
\(\ket{\eta}\) is an IDF's state vector. Note that for the
interference phenomena in \(K^{t}\), or more generally, in the
Hushimi representation of state vectors~\cite{HushimiRep}, we have an
established semiclassical interpretation~\cite{Adachi}, which will be
employed below.  

In order to investigate the influence of IDF's quantum oscillations 
on the EDF in a purified manner, {\em the two-state linear curve
crossing model for an infinitely heavy particle\/} (for short, the heavy
particle model) is studied. The IDF of this model is a two-level
system, which is described by Pauli matrices. 
The heavy particle model is described by the Hamiltonian that does not 
have EDF's kinetic term
\begin{equation}
  \label{eq:ModOne}
  \hat{H} \equiv \hat{V} (\hat{q}) = 
    - \hbar \pauli_z F\hat{q}  + \hbar \pauli_x J.
\end{equation}
The Hamiltonian is scaled by the Planck constant \(\hbar\) in order to 
retain the independence of IDF's time scales from \(\hbar\).
During a time evolution, the position of the EDF is an invariant.
On the contrary, the momentum of the EDF is excited: In the absence of 
\(J\), the EDF feel the force \(+F\) (\(-F\)) when the state of the
IDF is \(\ket{\uparrow}\) (\(\ket{\downarrow}\)).
The transition matrix element \(J\) induces the quantum oscillation of 
the IDF between \(\ket{\uparrow}\) and \(\ket{\downarrow}\). Hence we
lose the classical picture of the force acting on the EDF.

In order to treat only the EDF semiclassically, an effective action
for the EDF is introduced:
\begin{equation}
  \label{eq:DefSEff}  
  S^{\rm eff}(q) = - i \hbar \ln Z(q),
\end{equation}
where the ``influence functional''~\cite{FN:FeynmanPath}
(cf. Ref.~\cite{FeynmanVernon}) is defined as
\begin{equation}
  Z(q) = \bra{\eta''}{e^{-i \hat{V} (q) t / \hbar}}\ket{\eta'}.
\end{equation}
By employing \(S^{\rm eff}\) as a ``classical action'', the Feynman
kernel \(K^{t}\) is expressed as a coherent-state path
integral~\cite{DaubechiesKlauder} of the  
EDF. The semiclassical theory employed here is the stationary phase
evaluation of the coherent-state path integral~\cite{Klauder}.
A stationary phase point is specified by the complex classical
trajectory \((\bar{q}, \bar{p})\) that obeys the following 
Hamilton equation~(symplectic mapping)
\begin{equation}
  \label{eq:HamEq}
  \bar{p}'' = \bar{p}' + \partial S^{\rm eff}(\bar{q}') / \partial q, \quad
  \bar{q}'' = \bar{q}',
\end{equation}
where single- and double-primed quantities correspond to the initial and the
final times, respectively.
Furthermore, the ``entrance label'' \((q', p')\) and the ``exit
label'' \((q'', p'')\) of \(K^{t}\) specify the boundary condition of 
\((\bar{q}, \bar{p})\)~\cite{Klauder}
\begin{equation}
  \label{eq:KBC}
  P'  = (p'  - i q')  / \sqrt{2}, \quad
  Q'' = (q'' - i p'') / \sqrt{2},
\end{equation}
where \((Q,P)\) are defined by the linear canonical transformation
\(P = (\bar{p} - i \bar{q}) / \sqrt{2}\) and 
\(Q = (\bar{q} - i \bar{p}) / \sqrt{2}\)~\cite{Kramer}.
A solution of (\ref{eq:HamEq}) and (\ref{eq:KBC}) can be specified by the
initial value of \(Q'\). The corresponding semiclassical amplitude is 
\(E(Q') \exp \{i F(Q') / \hbar \}\), where \(E\) and \(F\) are the amplitude
and the ``action'', respectively.
Klauder expected that the semiclassical Feynman kernel 
\(K^{t}_{\rm SC}\) has always only single contribution of the
semiclassical amplitude~\cite{Klauder}. Actually, in a very short time
scale, this is the case. However, Adachi showed that in general
\(K^{t}_{\rm SC}\) has multiple contributions of the semiclassical
amplitudes in order to describe quantum interference
phenomena~\cite{Adachi}. 

\begin{figure}
  \hbox{%
    \epsfxsize = 0.45\textwidth \epsfbox{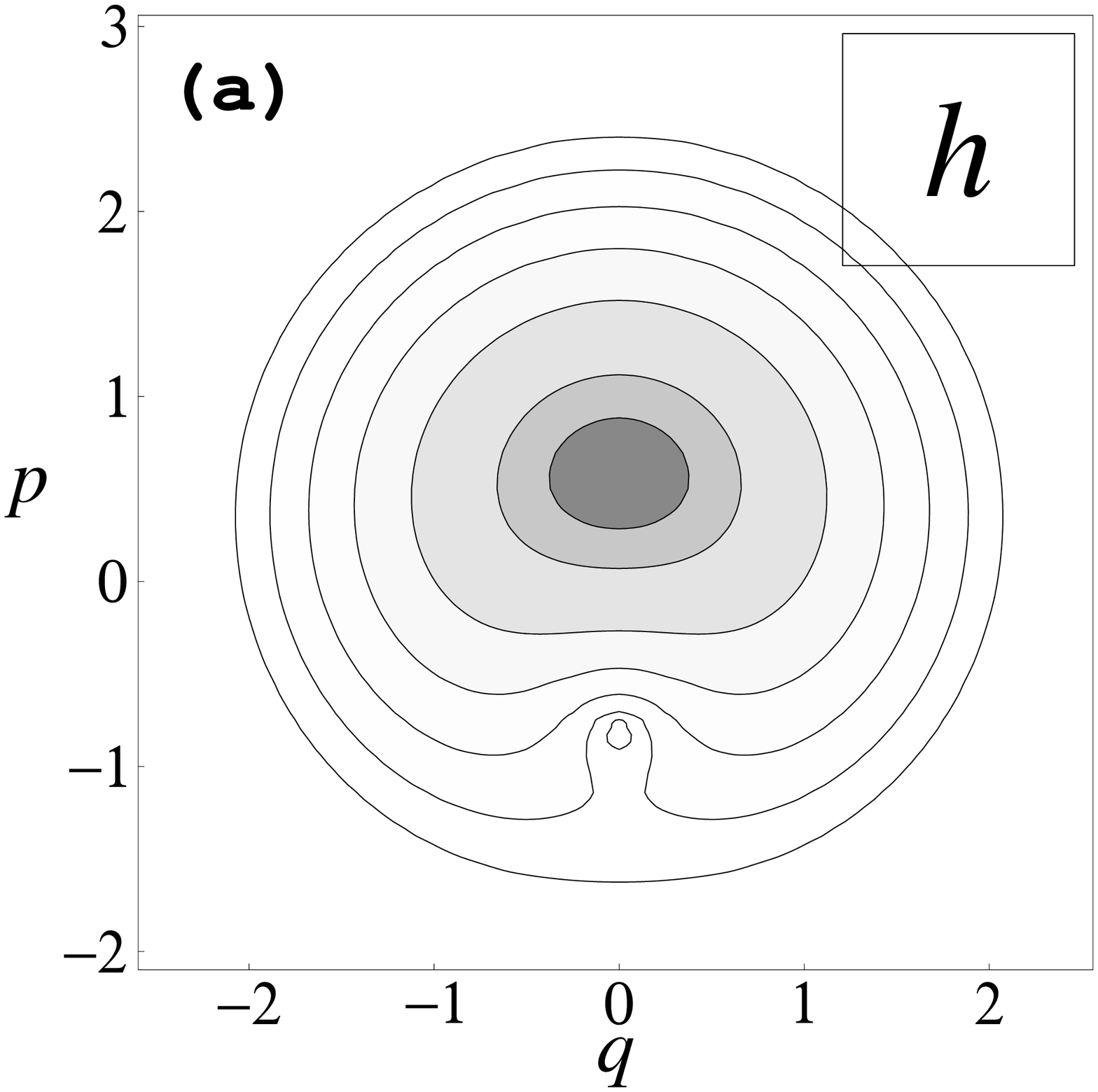} \quad
    \epsfxsize = 0.45\textwidth \epsfbox{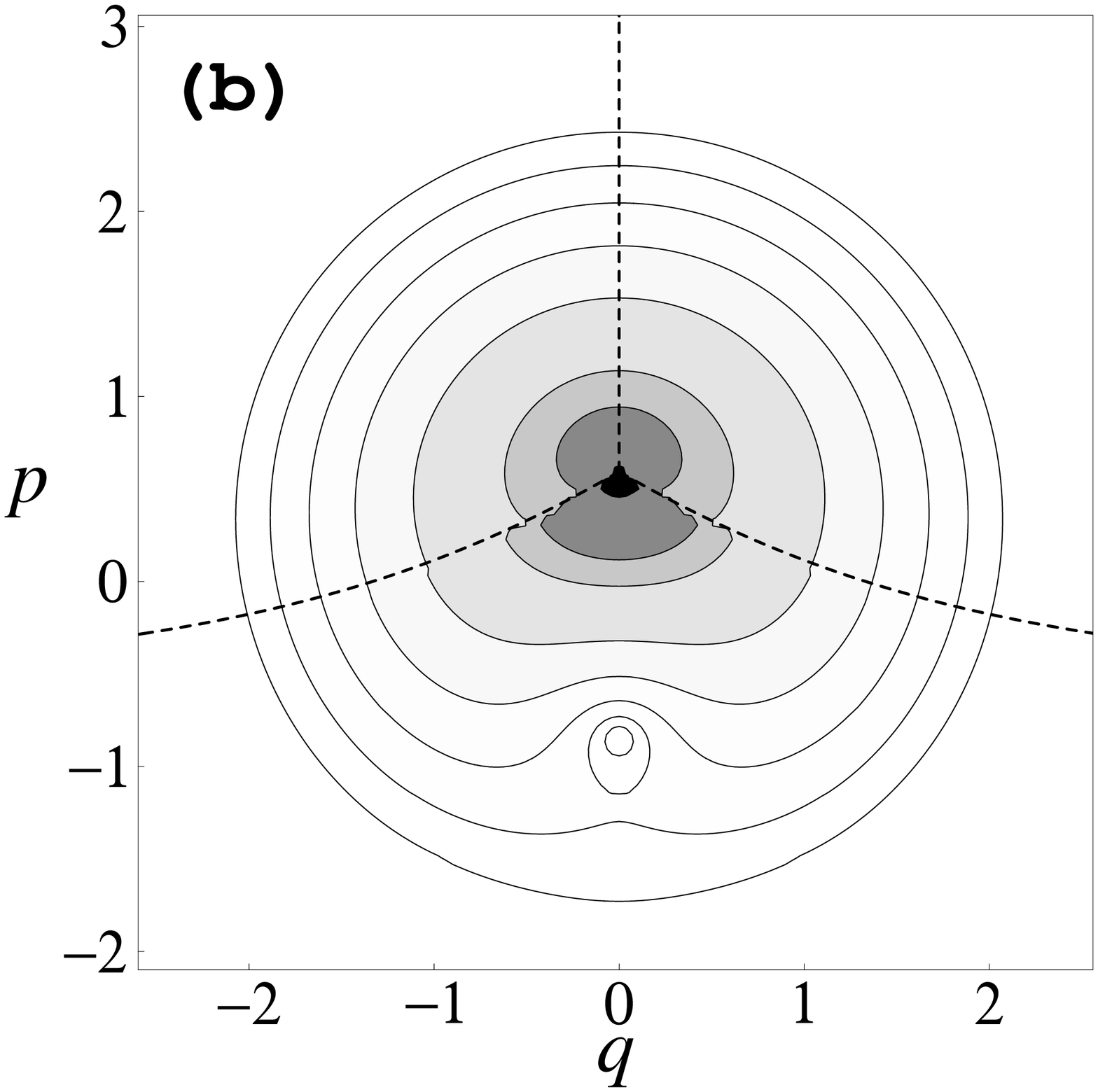}}
  \caption{Contour plots of a Hushimi function
    $|K^{t} (q p \uparrow; q'p' \uparrow)|^2$ 
    calculated by (a) the quantum theory and (b) the semiclassical theory.
    Parameters are $\hbar = h/(2\pi) = 0.25$ (indicated by a box in
    (a)), $F = 1.0$, $J = 0.75$, $(q',p') = (0, 0)$ and
    $t = 1.5$. At the same time, the Stokes lines are indicated
    by dashed lines in (b).}
  \label{fig:HusOne}
\end{figure}

In Fig.~\ref{fig:HusOne}, the ``exact'' evaluation and the
semiclassical evaluation of the Hushimi function \(|K^{t}|^2\) are shown: 
the semiclassical theory reproduces the exact Hushimi
function well. In the following, the semiclassical theory elucidates
the structure of the Hushimi function. Since the initial
condition of the IDF is  \(\ket{\eta'} = \ket{\uparrow}\), the center
of the amplitude of the EDF moves upward in the phase space due to the
diagonal element \(-\hbar F q\) of Eq.~(\ref{eq:ModOne}).  
At the same time, there is a zero of the Hushimi function at 
\((q, p) \sim (0, -0.8)\). Namely, we encounter 
{\it a quantum interference phenomenon}.

In ``single'' component systems, quantum interference patterns in the
Hushimi representation have intimate correspondence with phase space
caustics (PSCs)~\cite{Adachi}, which are the zero points of the Jacobian
\(\partial Q''(Q')/\partial Q'\). At a PSC, the semiclassical amplitude 
\(E = (\partial Q''/\partial Q')^{(-1/2)}\)~\cite{RubinKlauder} 
diverges. Furthermore, around the PSC, a pair of semiclassical
trajectories that are specified by two values of \(Q'\) appear for one
value of the exit label \((q'',p'')\).  
The contributions of the resultant multiple semiclassical amplitudes
to the Feynman kernel are controlled by the Stokes
phenomena~\cite{Stokes,Adachi}:  In one region of the \(Q'\)-plane, one of
the semiclassical amplitudes 
is ``unphysical'' so must be excluded. The boundary of the unphysical
region in the \(Q'\)-plane is the Stokes lines; In the other region,
the two semiclassical amplitudes contribute at once. 
Accordingly, a destructive interference pattern emerges in the Hushimi
function~\cite{Adachi}. Similarly as in the case of single component
systems, a PSC produced the interference pattern in Fig.~\ref{fig:HusOne}. 

\begin{figure}
  \epsfxsize = 0.45\textwidth \epsfbox{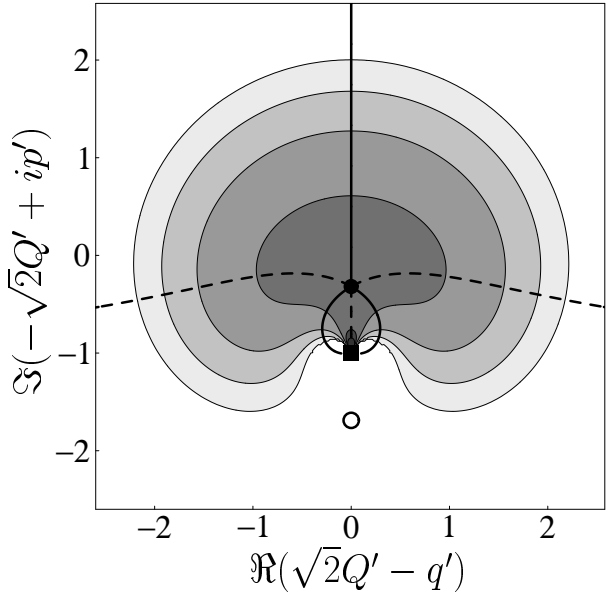}  
  \caption{Contour plot of $\Im F(Q')$, the imaginary part of the action
    for $K^{t}(q''p''\uparrow;q' p' \uparrow)$.
    The logarithmic divergent point of the effective action is
    indicated by $\blacksquare$. The corresponding pair of the PSC
    is indicated by $\bullet$ and $\circ$.
    As is explained in the main text, it is safe to ignore $\circ$.
    %
    Accordingly, only for $\bullet$, the Stokes lines (bold),
    which are part of the pre-image of the Stokes lines 
    in the $Q''$-plane, are drawn; Besides, the bold-dashed lines are
    the rest of the pre-image.
    The unphysical region is enclosed by the two Stokes lines that
    connect $\bullet$ and $\blacksquare$.
    Parameters are the same as in Fig.~\protect\ref{fig:HusOne}.}
  \label{fig:HPImF}
\end{figure}

The dynamical origin of PSCs is the ``folding'' dynamics due to
nonlinearity, especially the chaos, for the case of single component
systems~\cite{Adachi}.
For multi-component systems, we encounter a brand-new source of PSCs,
the logarithmic divergences of \(S^{\rm eff}\) due to the zeros of the
influence functional \(Z\). Indeed the interference pattern in
Fig.~\ref{fig:HusOne}~(b) is due to such a PSC. 
The general feature around a zero point of \(Z\) is explained
by expanding \(Z\) around the zero point~\cite{FN:ExpandZ,ATFuture}.
A zero of \(Z\) produces a pair of PSC.
One of the PSC can be safely ignored, since the value of \(\Im F\) is
too large to contribute to the Feynman kernel. For the other PSC, we
must treat the Stokes phenomena. 
The Stokes lines
in the \(Q'\)-plane are shown in Fig.~\ref{fig:HPImF}. 
According to the shape of these lines (looks like the upside-down
``Venus'' mark), I call the PSC that is caused by zeros of \(Z\), 
{\em v-PSC},  in order to distinguish the conventional ones, which is
called {\em a-PSC}, in the following argument. 

The coherent quantum oscillation of the IDF produce the zeros of
\(Z\), as is explained below. With a given complex classical
trajectory \(\bar{q}\), the IDF is evolved by \(\hat{V}(\bar{q})\).
Hence, during a time evolution, the value of \(Z\) oscillates due to 
the quantum oscillation of the IDF.
In particular, \(Z = 0\) holds when the state vector of
the IDF is orthogonal to \(\ket{\eta''}\), which is specified by the
exit label of \(Z\). With a fixed value of \(t\), we also encounter
the zeros of \(Z\) in varying \(\bar{q}\).

\paragraph*{The chaotic dynamics of the ``classical'' subsystem
  destroys the coherent quantum oscillation of the quantum subsystem}
%
We saw above that the effect of IDF's quantum oscillation appears as
v-PSCs at EDF's semiclassical dynamics. 
In turn, I will examine the effect of EDF's dynamics, especially
the chaotic dynamics, on v-PSCs by employing 
{\it the spin-kicked rotor\/} (the kicked rotor for a
spin-\(\frac{1}{2}\) particle)~\cite{Scharf,AT96}.
The spin-kicked rotor is composed of a rotor as an EDF and a two-level
system as an IDF and is described by the following Hamiltonian: 
\begin{equation}
  \label{eq:ModTwo}
  \hat{H}(t) \equiv
    T(\hat{p}) \hat{1}_{\rm I} + \hat{V}(\hat{q}) \sum_n \delta (t-n),
\end{equation}
where \(\hat{1}_{\rm I}\) is the identity operator of the IDF, 
\(T(p) \equiv p^2/2\) and 
\(\hat{V}(q) \equiv \hat{1}_{\rm I} K \cos q 
     + \hbar \pauli_z \delta K \cos q + \hbar \pauli_x J\).
This model is an extension of the standard
mapping~\cite{StandardMapping} to multi-component systems.
Corresponding to the periodically time-dependent Hamiltonian
\(\hat{H}(t)\), we have a Floquet operator
\(\hat{U} = 
        \exp [-i T(\hat{p}) / \hbar] 
        \exp [-i \hat{V}(\hat{q}) / \hbar] \).

\begin{figure}
    \hbox{%
      \epsfxsize=0.45\textwidth \epsfbox{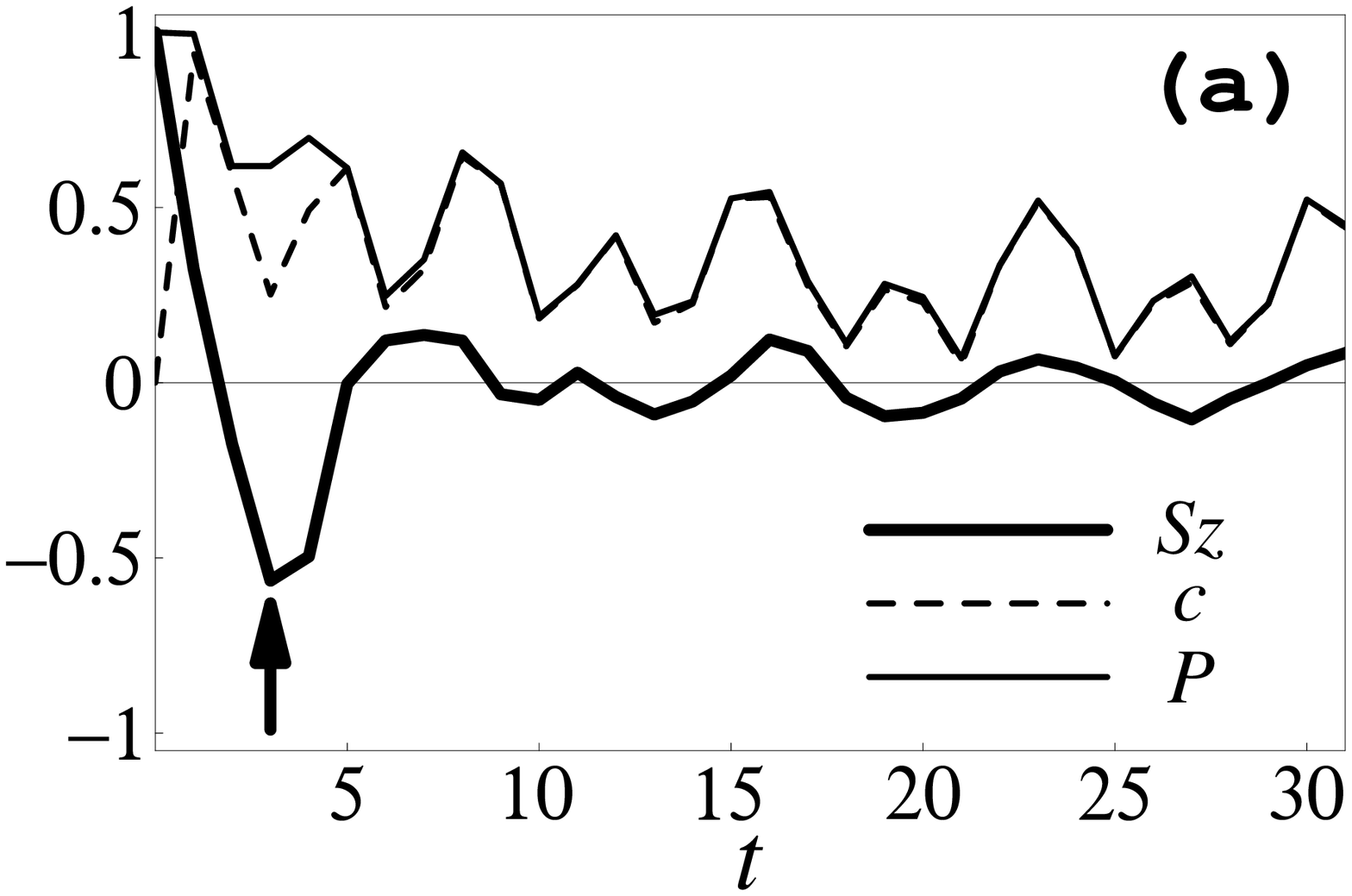} \quad 
      \epsfxsize=0.45\textwidth \epsfbox{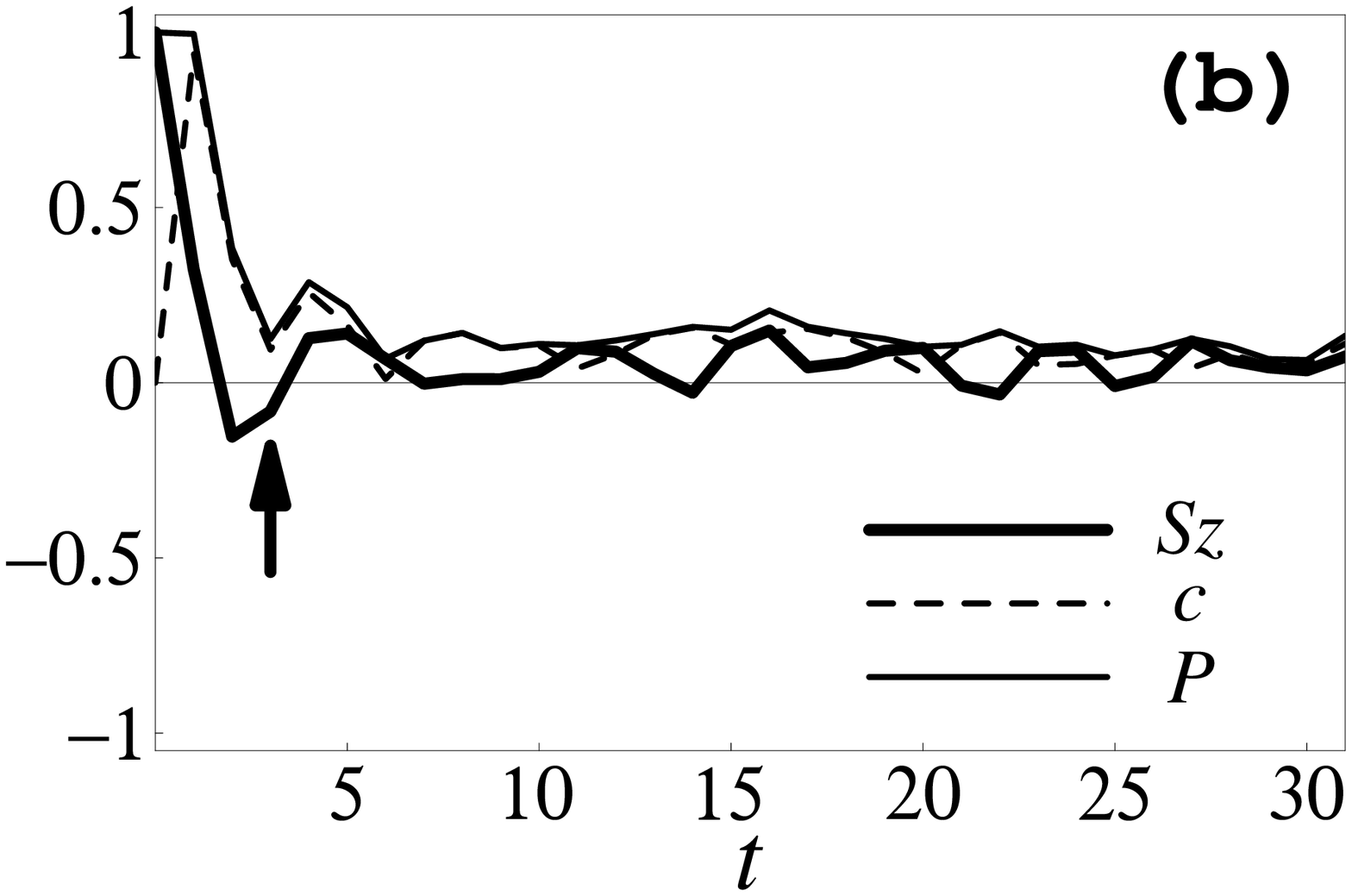}}
  \caption{
    Time evolutions of 
    $s_z \equiv \protect\qE{\hat{\sigma}_z}$,
    $c \equiv \protect\sqrt{\protect\qE{\hat{\sigma}_x}^2 
      + \protect\qE{\hat{\sigma}_y}^2}$ and
    $P \equiv \protect\sqrt{\protect\qE{\hat{\sigma}_x}^2 
      + \protect\qE{\hat{\sigma}_y}^2
      + \protect\qE{\hat{\sigma}_z}^2}$~\protect\cite{AT96}
    of (a) the regular case ($K = 0.4$) and (b) the chaotic case 
    ($K = 2.4$).
    The initial condition is $\protect\ket{q'p', \uparrow}$ with 
    $(q',p') = (0.0, 1.5)$.
    Parameters are $\hbar = 0.25$, $\delta K = 1.0$, and $J = 0.75$.
    In the long time evolution, the oscillation of these quantities,
    especially $c$, are suppressed in the chaotic case (b).}
  \label{fig:TLSEvol}
\end{figure}

In Fig.~\ref{fig:TLSEvol}, the time evolutions of the regular 
(\(K = 0.4\)) and the chaotic (\(K = 2.4\)) cases of quantities
concerning to the IDF are shown. At the third step (indicated by
arrows in the figures), the quantum oscillation continues in the
regular case, but decays in the chaotic case. These different short
time behaviors, which will be explained below by the semiclassical
theory, determine the long-time behaviors of the system, i.e.\ the
continuation of the coherent oscillation in the regular case
(Fig.~\ref{fig:TLSEvol}~(a)) and the destruction of the oscillation in
the chaotic case (Fig.~\ref{fig:TLSEvol}~(b)).  

In order to give a semiclassical interpretation of the phenomena
mentioned above, the ``full'' Feynman kernel 
\(\bra{q'' p'', \eta''}{\hat{U}^N}\ket{q' p', \eta'}\)
is studied by ``decomposing'' it by a sequence of the IDF's states
\(\{\eta_n\}_{n = 1}^{N-1}\)
\begin{eqnarray}
  \label{eq:DefDecoKer}
  &{}& \DecoKer (q''p''\eta'';\{\eta_n\}_{n = 1}^{N-1};q'p'\eta') 
            \nonumber \\
  &\equiv& \bra{q'' p''} (\bra{\eta_N}{\hat{U}}\ket{\eta_{N-1}}
    \ldots \bra{\eta_1}{\hat{U}}\ket{\eta_0} ) \ket{q' p'},
\end{eqnarray}
where \(\eta_N = \eta''\) and \(\eta_0 = \eta'\).
The decomposition of the kernel facilitate the semiclassical analysis:
For the full kernel, we have to solve the EDF's equation of motion that is
nonlocal in time~\cite{Pechukas}; On the contrary, for \(\DecoKer\), 
the equation of motion is local in time.
In evaluating \(\DecoKer\) by a stationary phase method concerning to the 
EDF, similar to the analysis of the heavy particle model
(\ref{eq:ModOne}), let us introduce an effective potential
\(V^{\rm eff}_n (q) \equiv i \hbar \ln Z_n (q)\),
where  \(Z_n (q) \equiv
        \bra{\eta_n}{\exp [-i \hat{V}(q) / \hbar]}\ket{\eta_{n-1}}\).
The interaction term of the EDF's effective action is 
\(S^{\rm eff}_{\rm int} = - \sum_{n=1}^{N} V^{\rm eff}_n\).
Accordingly, the classical equation of motion for the complex
classical trajectory \(\{(\bar{q}_n, \bar{p}_n)\}_{n}\) is  
\begin{equation}
  \label{eq:HamEqTwo}
  \bar{p}_n = \bar{p}_{n-1} 
              - \partial V^{\rm eff}_n (\bar{q}_{n-1}) / \partial q,
  \quad
  \bar{q}_n = \bar{q}_{n-1} + \bar{p}_n.
\end{equation}
At the same time, Klauder's boundary condition (\ref{eq:KBC}) is
imposed on the complex classical trajectory.

In the semiclassical study of \(\DecoKer\), v-PSCs appear with the
similar mechanism mentioned above. Furthermore, we have to discuss the 
reconstruction of the full kernel from \(\DecoKer\). However,
concerning to the quantum oscillation of the IDF in the short time
scale, the reconstruction of the full kernel plays no particular role,
as is confirmed from the numerical observations. Hence, I report 
only the semiclassical study of \(\DecoKer\).

\begin{figure}
    \hbox{%
      \epsfxsize = 0.35\textwidth \epsfbox{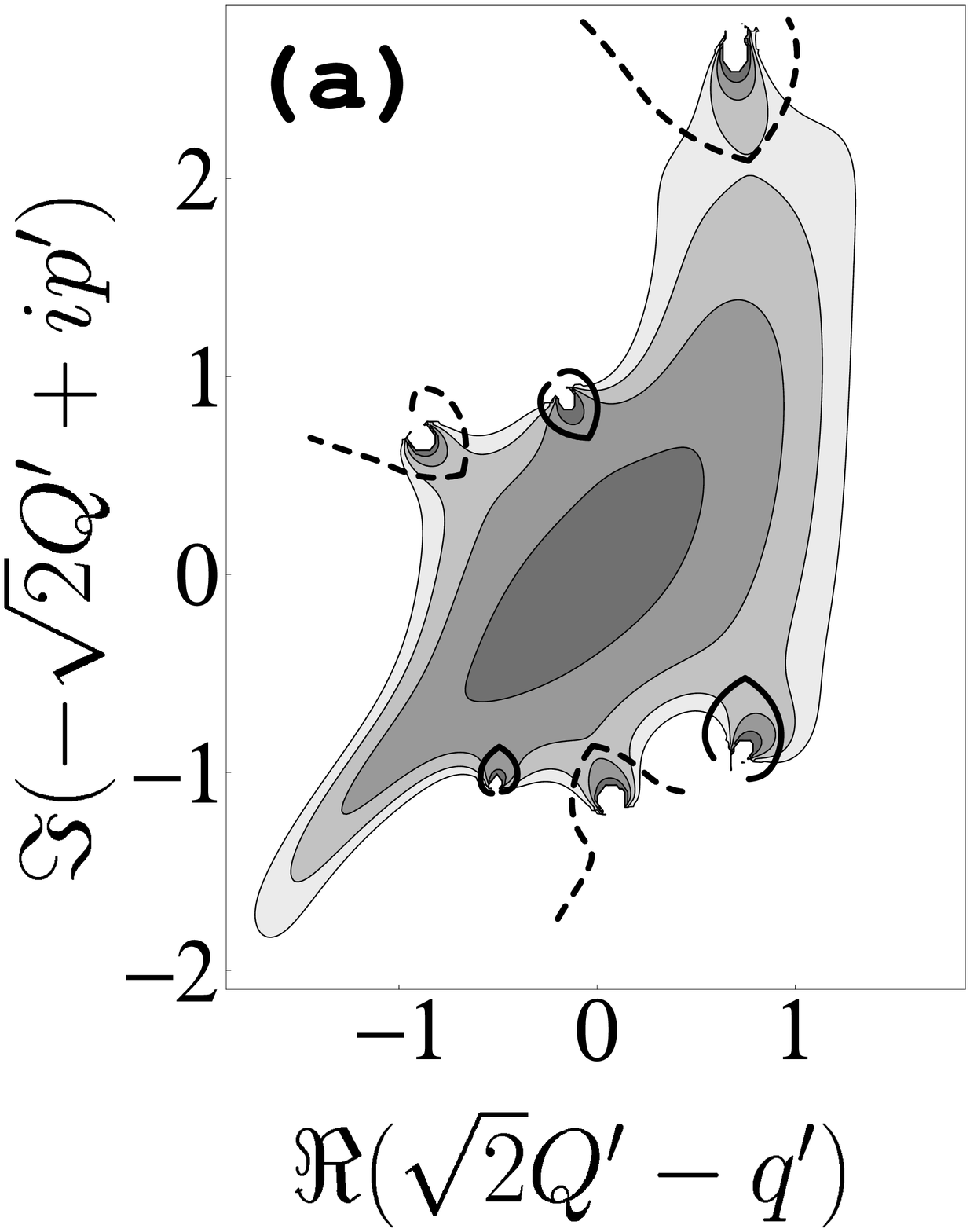} \quad
      \epsfxsize = 0.55\textwidth \epsfbox{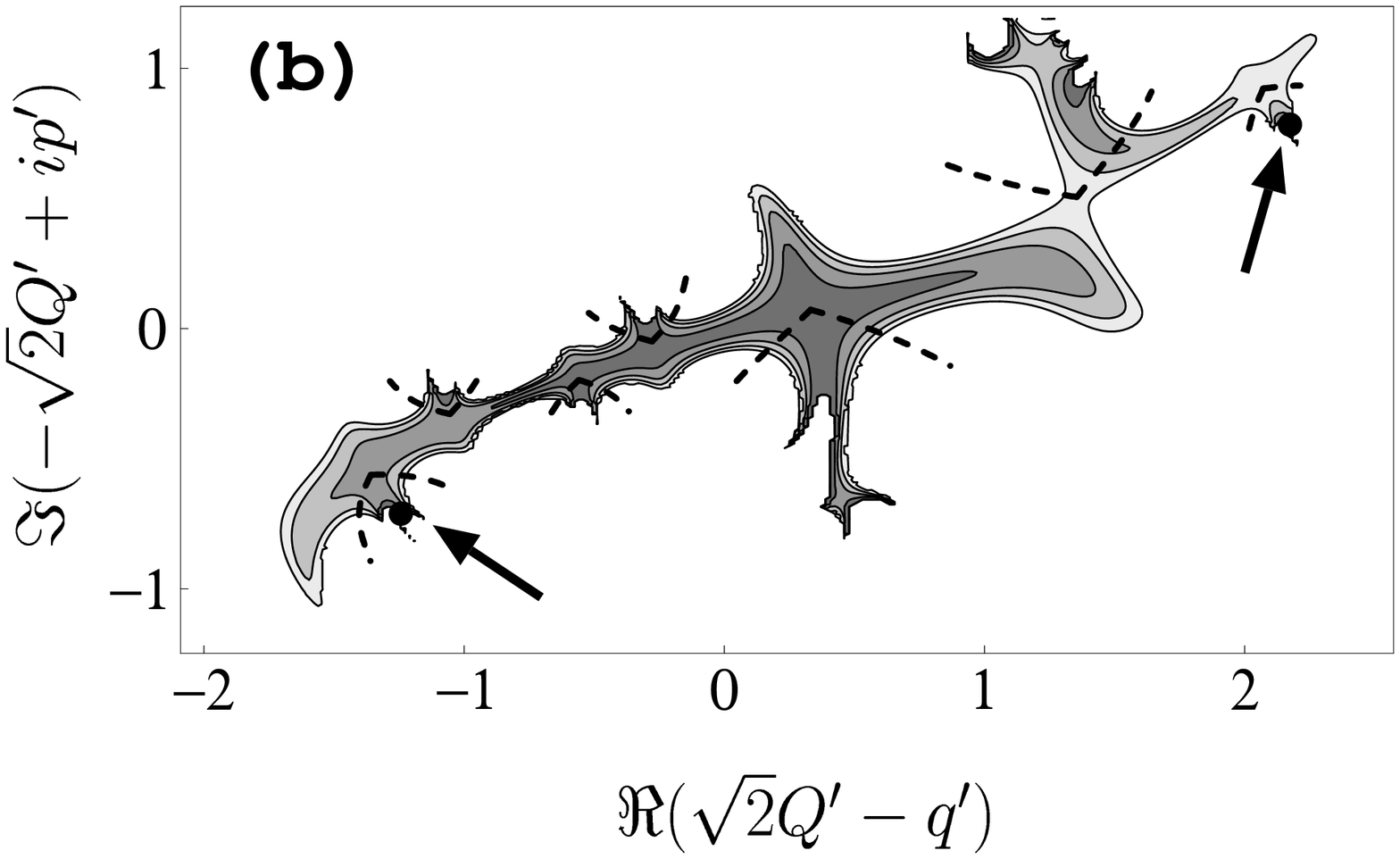}}
  \caption{
    Contour plot of $\Im F(Q')$ in the domain
    $D$~(\protect\ref{eq:DefDomain}) ($(\Im F)_{\rm cutoff} = 1.151$) of
    $K^{3}_{\rm d} (q''p''\uparrow; \{\uparrow ,\uparrow\};q'p'\uparrow)$
    for (a) the regular ($K = 0.4$) and (b) the chaotic ($K = 2.4$) 
    cases. The Stokes lines that start from v-PSCs (solid lines)
    and a-PSCs (dashed lines) are also shown. Parameters are the
    same as in Fig.~\protect\ref{fig:TLSEvol}.
    In order to treat the Stokes phenomena, the region
    that are enclosed by the Stokes lines must be regarded as
    unphysical (non-contributing) region. Note that in the chaotic 
    case (b), all of v-PSCs (indicated by $\bullet$ and pointed by
    arrows) in $D$ are in the unphysical region produced by a-PSCs.
    }
  \label{fig:SKImF}
\end{figure}

In evaluating \(\DecoKer\) semiclassically, a ``physical'' region
\(D\) on the \(Q'\)-plane (i.e.\ iinitial points of the complex
trajectory) is introduced~\cite{Adachi}:
\begin{equation}
  \label{eq:DefDomain}
  D = \{Q' | \Im F(Q') \leq (\Im F)_{\rm cutoff} \},
\end{equation}
where \(F\) is the classical action for \(\DecoKer\).
If \(Q'\) is out of the region \(D\), the corresponding semiclassical
amplitudes are too small to contribute to \(\DecoKer\). Hence it is
enough to count the contribution only from \(D\) for the semiclassical
Feynman kernel. 
For single component systems, a perturbation analysis shows that
the chaotic dynamics makes \(D\) contract exponentially fast around the
classically realizable trajectory, whose stability exponent determines 
the rate of the contraction~\cite{ATFuture}. 
Although the multi-component systems do not have any classically realizable
trajectory, the similar contraction of \(D\) was observed in the
numerical experiment in the chaotic case (Fig.~\ref{fig:SKImF}~(b)).
Furthermore, I observed that the contraction of \(D\) have different
influences on two kinds of PSCs;
On one hand, a-PSCs produced by the chaotic
dynamics catch up the contraction of \(D\).
Accordingly, a-PSCs have significant influence on the Feynman kernel,
similarly to the single-component case;
On the other hand, v-PSCs fail to catch up \(D\) due to the
contraction of \(D\), though v-PSCs have strong influence in the
regular case (Fig.~\ref{fig:SKImF}~(a)).
Due to the contraction of \(D\), the unphysical region produced by
v-PSCs become smaller. Furthermore, some v-PSCs move into the
unphysical region produced by a-PSCs (Fig.~\ref{fig:SKImF}~(b)).
Consequently, contributions of v-PSCs, which implies the coherence of
the IDF's quantum oscillations, to the Feynman kernel is suppressed by
the EDF's chaotic dynamics and thus the IDF's quantum oscillations,
which have strong correspondence to the cause of v-PSCs, becomes
incoherent (Fig.~\ref{fig:TLSEvol}~(b)).

\paragraph*{Discussion}
%
The semiclassical method employed for the analysis of the spin-kicked 
rotor in the latter part of this paper has only a limited
applicability: It is useful only for short time steps of quantum
mapping systems and difficult to apply for quantum flow
systems. Hence, we have to develop the semiclassical method that
really works with the more general quantum-``classical'' coupling
systems. However, it is plausible that similar mechanisms obtained in
this letter generally appear concerning to the interactions between
the quantum oscillation of the quantum subsystem and the dynamics of
the classical subsystem.

\acknowledgments
%
I wish to thank Professor K.~Kitahara and Dr. S. Adachi for
encouragement. I am grateful to Dr. M. Wilkinson for the hospitality
during my stay at the University of Strathclyde, where this manuscript
is partially prepared: The EPSRC (grant no.~GR/L02302) supported my
stay there. 
I also thank Dr.~M.~Murao for discussing on quantum coherence. 
Numerical computations are carried out at the Kitahara-Hara group at
the Tokyo Institute of Technology, and at the Yukawa Institute for
Theoretical Physics. I wish to thank Professor K. Kitahara, 
Professor T. Hara and Dr. S. Adachi for providing the computational
environment at the Tokyo Institute of Technology.



\onecolumn
\input epsf.sty

\begin{thebibliography}{99}
\bibitem[*]{YRF}{Yukawa research fellow}
\bibitem[\dagger]{EMAIL}{E-mail: {\tt atanaka@cm.ph.tsukuba.ac.jp}}
\bibitem[\ddagger]{PresentAddress}
  {Present address: Institute of Physics, University of Tsukuba, 
    Tsukuba, Ibaraki \mbox{305-0006}, Japan} 
\bibitem{dEspagnat}
  {See, e.g., B. d'Espagnat, 
    {\it Conceptual foundations of quantum mechanics} 
    (W. A. Benjamin, Massa\-chu\-setts, 1976)}
\bibitem{Molecule}
  {J. C. Tully and R. K. Preston, J. Chem. Phys. {\bf 55}, 562 (1971);
    W. H. Miller and T. F. George, J. Chem. Phys. {\bf 56}, 5637 (1972)}
\bibitem{Pechukas}{P. Pechukas, Phys. Rev. {\bf 181}, 174 (1969); 
    J. Cao, C. Minichino and G. A. Voth, J. Chem. Phys. {\bf 103}, 1391 (1995)}
\bibitem{StockThoss}
  {G. Stock and M. Thoss, Phys. Rev. Lett. {\bf 78}, 578 (1997)}
\bibitem{Schroedinger}
  {E. Schr\"odinger, Proc. Camb. Phil. Soc. {\bf 31}, 555 (1935)}
\bibitem{Gutzwiller}{M. C. Gutzwiller, 
    {\it Chaos in Classical and Quantum Mechanics} 
    (Springer-Verlag, New York, 1990)}
\bibitem{Glauber}{J. R. Glauber, Phys. Rev. {\bf 131}, 2766 (1963)}
\bibitem{HushimiRep}
  {K. Hushimi, Proc. Phys. Math. Soc. Jpn. {\bf 22}, 264 (1940); 
    K. Takahashi and N. Sait\^o, Phys. Rev. Lett. {\bf 55}, 645 (1985)} 
\bibitem{Adachi}{S. Adachi, Ann. Phys. {\bf 195}, 45 (1989)}
\bibitem{FN:FeynmanPath}{Since the time evolution under consideration
    can be regarded as a ``unit time'' evolution of a quantum mapping
    system, the Feynman path is a variable \(q\) rather than a
    function of time.}
\bibitem{FeynmanVernon}
  {R. P. Feynman and F. L. Vernon, Ann. Phys. {\bf 24}, 118 (1963)}
\bibitem{DaubechiesKlauder}
  {I. Daubechies and J. R. Klauder, J. Math. Phys. {\bf 26}, 2239 (1985)}
\bibitem{Klauder}
  {J. R. Klauder, in {\it Path Integrals}, Proceedings of the NATO
    Advanced Summer Institute, edited by G. J. Papadopoulos and
    J. T. Devreese, (Plenum, New York, 1978), p. 5;
    J. R. Klauder, in {\em Random Media}, edited by G. Papanicolauou,
    (Springer-Verlag, New York 1987), p. 163}
\bibitem{Kramer}{P. Kramer, M. Moshinsky and T. H. Seligman,
    Group Theory and Its Applications {\bf III}, 249 (1975)}
\bibitem{RubinKlauder}
  {A. Rubin and J. R. Klauder, Ann. Phys. {\bf 241}, 212 (1995)}
\bibitem{Stokes}{G. G. Stokes, Trans. Camb. Phil. Soc. {\bf 10}, 106 (1864)}
\bibitem{FN:ExpandZ}{Around the zero of \(Z\), we can regard \(Z\) as
    a prefactor of the oscillatory integral (i.e., exclude \(\ln Z\)
    term from the effective action). However, such treatment prevent
    our understanding of the stationary points in a uniform manner in
    the \(Q'\)-plane.} 
\bibitem{ATFuture}{Atushi Tanaka, in preparation}
\bibitem{Scharf}{R. Scharf, J. Phys. A: Math. Gen. {\bf 22}, 4223 (1989)}
\bibitem{AT96}{A. Tanaka, J. Phys. A: Math. Gen. {\bf 29}, 5475 (1996)}
\bibitem{StandardMapping}
  {B. V. Chirikov, Phys. Rep. {\bf 52}, 263 (1979); 
    G. Casati {\it et al.}, Lecture Notes in Physics {\bf 93}, 334 (1979)}
\end{thebibliography}
\end{document}